# A multi-channel wire gas electron multiplier

B. M. Ovchinnikov[1,*], V. V. Parusov[1], Yu. B. Ovchinnikov[2]

[1]Institute of Nuclear Research of Russian Academy of Sciences, Troitsk, Russia

[2]National Physical Laboratory, Teddington, Middlesex, TW11 0LW, UK

## Abstract

A novel and relatively simple method of production of electrodes for a multi-channel wire gas multiplier is developed. Two modifications of the multipliers have been tested: with a multiplication of electrons between two wire electrodes, MWGEM, and between a wire electrode and continuous anode, MWCAT.

For both MWGEM and MWCAT detectors, filled with neon under pressure of 760 Torr and irradiated by $\beta$-particles ($Ni^{63}$), the coefficient of proportional multiplication of electrons up to $10^4$ was obtained.

For the MWGEM detector irradiated by $\alpha$-particles ($Pu^{239}$), the coefficient of proportional multiplication of 300 was obtained.

It is observed, that in contrast to the GEM detectors, produced by perforation of a metal-clad plastic foil, in a MWGEM the discharges do not destroy its electrodes even for the potentials above the threshold of discharges.

The results on operation of the MWCAT filled with $Ar$, $Ar+CH_4$ and $Ar+1\% Xe$ are also presented.

## Introduction

A standard gas electron multiplier (GEM) [1] consists of a thin, metal-clad polymer foil, chemically pierced by a large number of holes. On application of a potential between the two metal facings of GEM, electrons, released by radiation in the gas at one side of the structure, drift into the holes, multiply and transfer to a collection region. The GEM detectors are used in high-energy physics for detection of different ionizing particles with a good spatial resolution. On the other hand GEM detectors suffer from one common problem: the streamers and sparks, which are developing at high gains of the detector, are unavoidably lead to its damage.

A great advantage of multi-channel gas electron multipliers (MWGEM) is that they are much less vulnerable to the unrecoverable destruction by streamers and sparks.

---

[*] e-mail: ovchin@inr.ru, tel: +8 (4967) 51 98 85

This paper presents the design and the results of tests of the two modifications of the multipliers, based on wire electrodes: with a multiplication of electrons between two wire electrodes, MWGEM, and between a wire electrode and continuous anode, MWCAT (an analogue of CAT [2]).

## MWGEM with two wire electrodes

The original design of our multi-channel wire gas electron multiplier (MWGEM) is described in [3] (Fig. 1). The MWGEM has openings of 0.5x0.5 mm$^2$, with 1.5 mm steps between them in two orthogonal directions. The gap between the electrodes is equal to 1 mm, while the total working area has a diameter of 20 mm. The gas multiplication of electrons in the MWGEM happens between the rectangular openings of the electrodes.

The previous tests of the MWGEM [3] were conducted only for the *Ar+20% $CO_2$* (760 Torr) gas mixture irradiated by $\alpha$-particles (*$Ra^{226}$*). In that case, the streamer discharges were observed for the gain factors of $K_{ampl}>10$, and the number of them was increasing with increasing of the voltage. As a result of a rapid increase of the number of streamers, no increase of the gain $K_{ampl}$ was observed above the voltages of 2.9-3.0 kV.

In this work we used the MWGEM with the same design of electrodes as in [3], but with better alignment of the wires, which has been achieved by using of a special template (Fig. 2) for the winding of the electrodes. The gap between the two electrodes of the MWGEM was equal to 1 mm. The design of the chamber for testing of the MWGEM detector is shown in Fig. 3. The gap between the cathode and the MWGEM was equal to 13 mm and a corresponding gap between the MWGEM and the anode was equal to 6 mm.

The chamber was filled with pure commercial neon at pressure of 760 Torr. The space in the gap between the cathode and the MWGEM was irradiated by $\alpha(Pu^{239})$ and $\beta(Ni^{63})$ particles.

The signals from the anode were amplified by a charge sensitive amplifier BUS 2-96.

At the irradiation of the chamber by *$\beta$-particles ($Ni^{63}$, ~100 $\beta$/sec)* the coefficient of proportional multiplication up to $8\cdot10^3$ was obtained (Fig. 4a). The streamer discharges were practically absent for the gains $K_{ampl}<8\cdot10^3$, while above that gain value the signals were consisting mostly from the streamer discharges.

The maximal coefficient of proportional multiplication of about 300 was obtained at the irradiation of the chamber by $\alpha$-particles *($Pu^{239}$)* (Fig. 4b).

# Multi-channel wire gas electron multiplier with wire cathode and continuous anode (MWCAT)

The prototype of GEM [1] and Micromegas [4] is a CAT [2], which consists of a net-cathode and a continuous anode, with a gas gap between them of 0.1-1.0 mm. The CAT detector with the same configuration of its winded cathode, as in the MWGEM, has been investigated. The tests of the MWCAT were conducted in a chamber (Fig. 5) with the gap between the cathode and the MWCAT equal to 13 mm.

The dependence of the gain $K_{ampl}$ = f($V_d$) has been obtained (Fig. 4c) at the filling of the chamber with a pure neon (760 Torr) and at irradiation by $\beta$-particles. In that case, the maximal coefficient of proportional amplification was about $1.5 \cdot 10^4$. There are only rare streamers have been observed at the gain of $K_{ampl} = 10^4$. For the higher gains the rapid increase of the number of streamer discharges was detected. Therefore, for the MWCAT detector, a sharp transition from the regime of proportional multiplication to the regime streamer discharges took place.

The results of measurements for a chamber filled with Penning mixture Ar+20% CH$_4$ (760 Torr) and irradiated by $\beta$-particles are presented in fig. 6. The maximal coefficient of proportional multiplication of about 100 was observed (curve a). In that case, the streamer discharges (curve b) have been observed for the whole range of proportional signals. The amplitude of streamer discharges stops to be increased at the voltage $V_d$ on a resistive divider equal to ~3 kV, that can be explained by the slowing of the increase of the potential difference between the MWCAT electrodes (Fig. 6c) due to increasing of the number of streamer discharges.

The shape of the streamer signal for *Ar+20% CH$_4$* gas mixture is presented in fig. 7. The large quantity of the streamers, as well as their long time duration, in this gas mixture can be explained by the formation of secondary avalanches, with duration of about 20 μs [5], which take place after each proportional signal.

Fig. 8 shows the results of the measurements performed for the chamber filled with pure commercial *Ar*. The maximal coefficient of proportional multiplication of electrons up to $10^3$ (curve a) has been observed, which is by an order of the magnitude greater, compared to the *Ar+20% CH$_4$* gas mixture. The streamer discharges were observed in the whole range of proportional signals, the amplitude of which was saturated at $V_d$ ~2 kV (curve b).

The results of measurements at filling of the chamber with mixture Ar+1% Xe (760 Torr) are presented in Fig. 9. In that case, the streamer discharges with average frequency of about 1 Hz were observed within the whole range of proportional signals.

## Conclusion

It is demonstrated, that the best operational parameters of MWGEM and MWCAT detectors are realized at filling of them with neon. It is shown also that a MWGEM is more reliable detector compared to a plastic GEM, because of much smaller probability of its damage by accidental discharges. If in a plastic GEM the discharges lead to producing of permanent short circuits between the electrodes of the detector, in MWGEM detectors this problem does not exist as well as the space between the electrodes is filled with gas only.

In a conclusion, the reliable MWGEM detectors can find applications not only in high-energy experimental physics and in X-ray astronomy, but also in medicine (X-ray examinations, positron tomography) and industrial non-destructive defectoscopy.

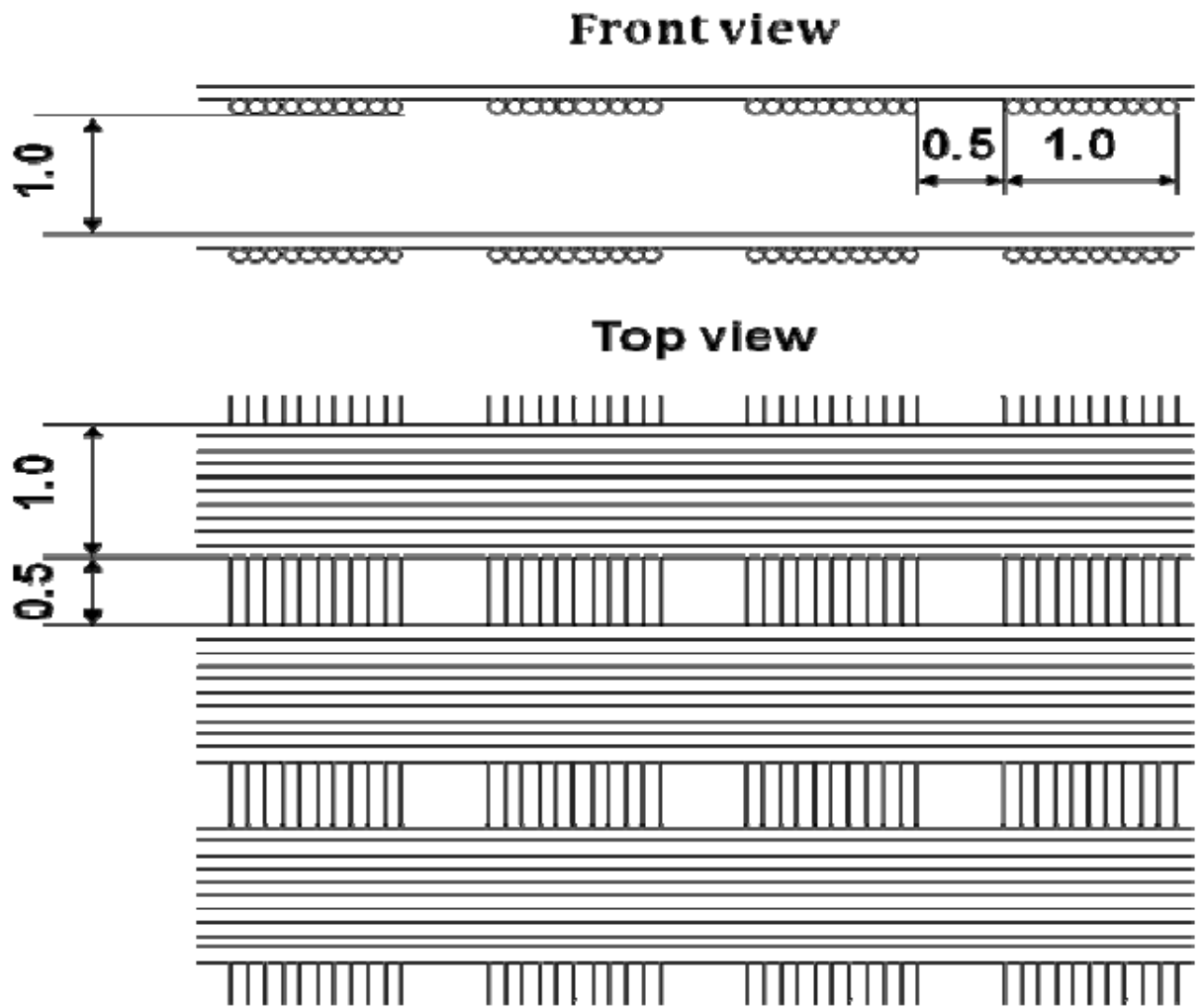

Fig. 1. MWGEM detector layout.

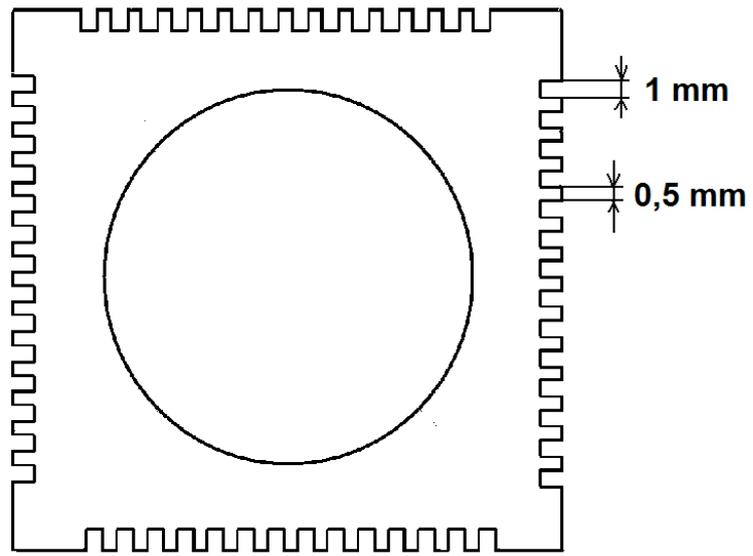

Fig. 2. Template for the winding of wire electrodes.

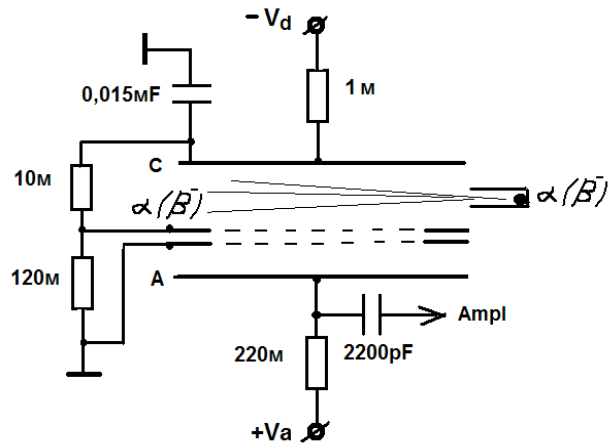

Fig. 3. The chamber for MWGEM investigations.

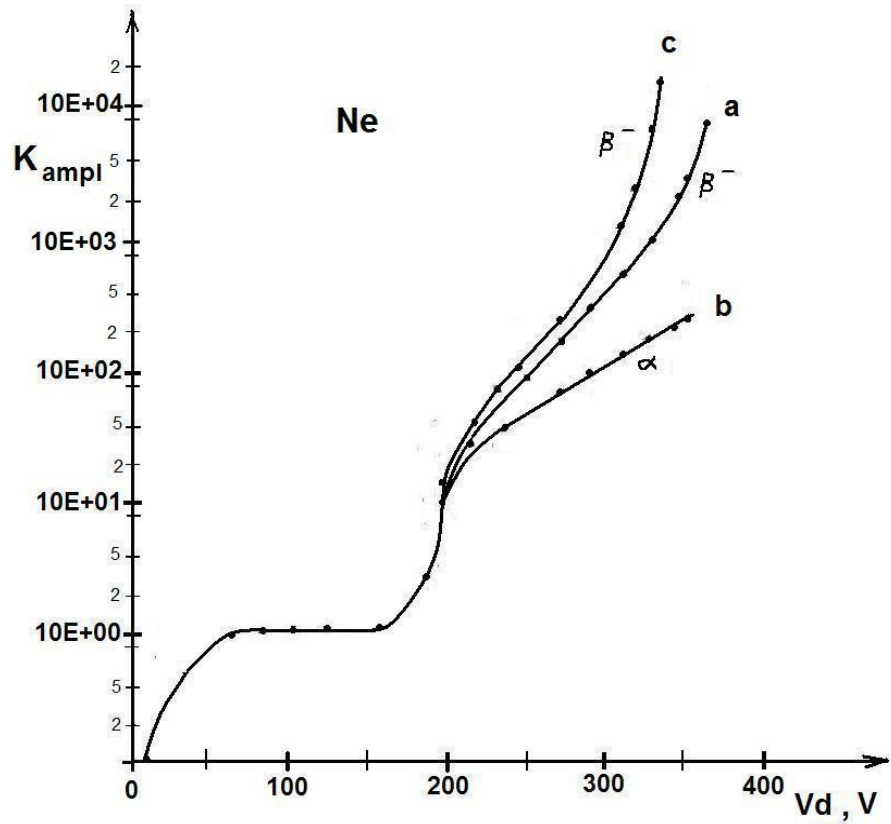

Fig. 4.  Curve (a): $K_{ampl} = f(V_d)$ for MWGEM at neon filling and at $\beta$-irradiation (Ni$^{63}$);

Curve (b): the same for $\alpha$-irradiation (Pu$^{239}$);

Curve (c): MWCAT at $\beta$-irradiation.

$V_d$ - voltage on resistive dividers of the chambers fig. 3 and fig. 5.

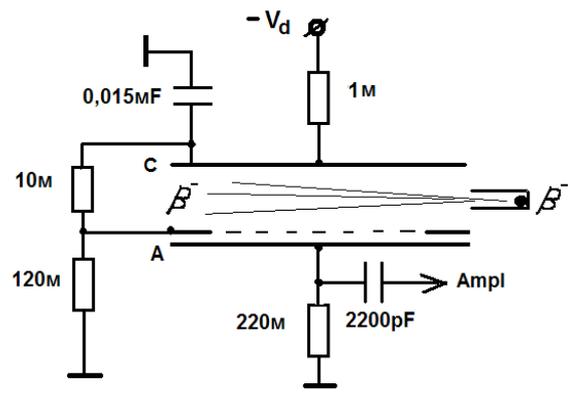

Fig. 5. The chamber for MWCAT investigations.

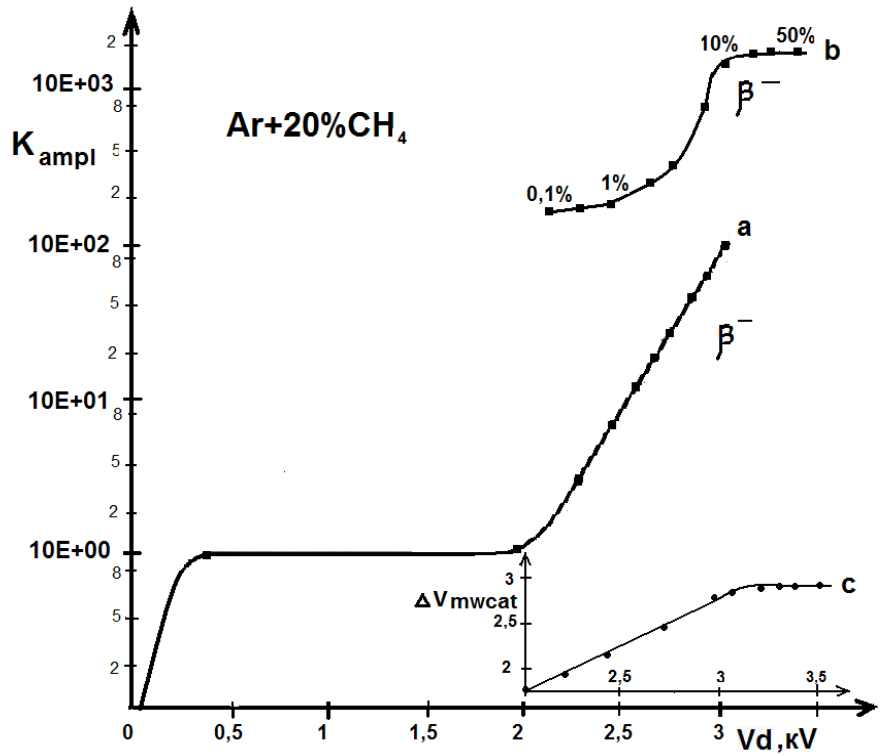

Fig. 6. Curve (a): The $K_{ampl} = f(V_d)$ of proportional signals for MWCAT at $Ar+20\% CH_4$ filling and $\beta$-irradiation;

Curve (b): $K_{ampl} = f(V_d)$ of streamer discharges for MWCAT at $Ar+20\% CH_4$ filling and $\beta$-irradiation;

Curve (c): voltage difference between MWCAT electrodes $\Delta V_{MWCAT}$ as a function of $V_d$.

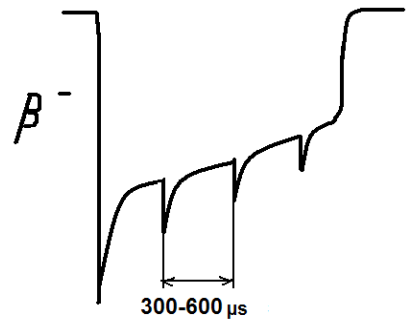

Fig. 7. Form of the streamer signal for *Ar +20 % CH₄* mixture and *β*- irradiation.

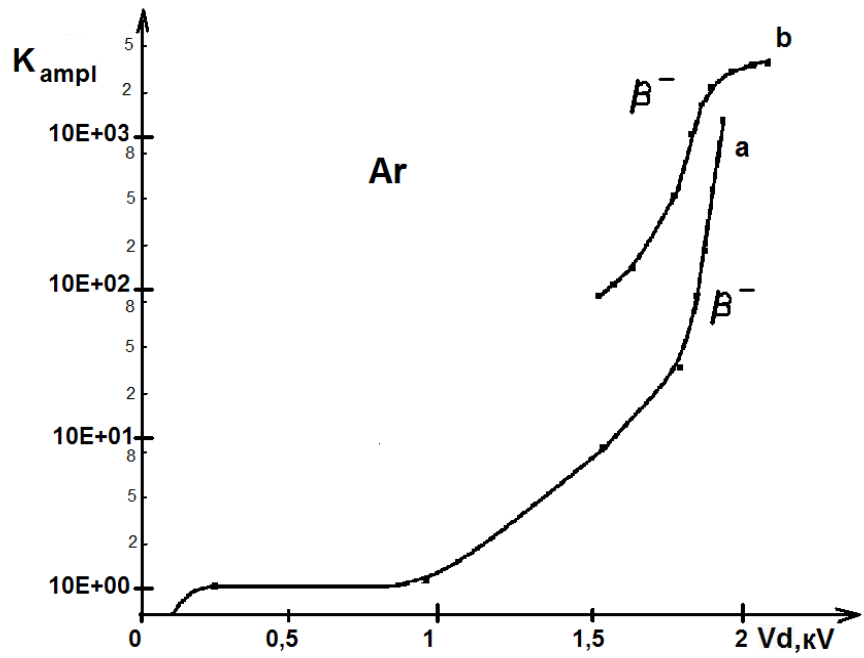

Fig. 8. Curve (a): $K_{ampl} = f(V_d)$ of proportional signals for MWCAT at $Ar$ – filling and $\beta$-irradiation;

Curve (b): $K_{ampl} = f(V_d)$ of streamer discharges for MWCAT at $Ar$ – filling and $\beta$-irradiation.

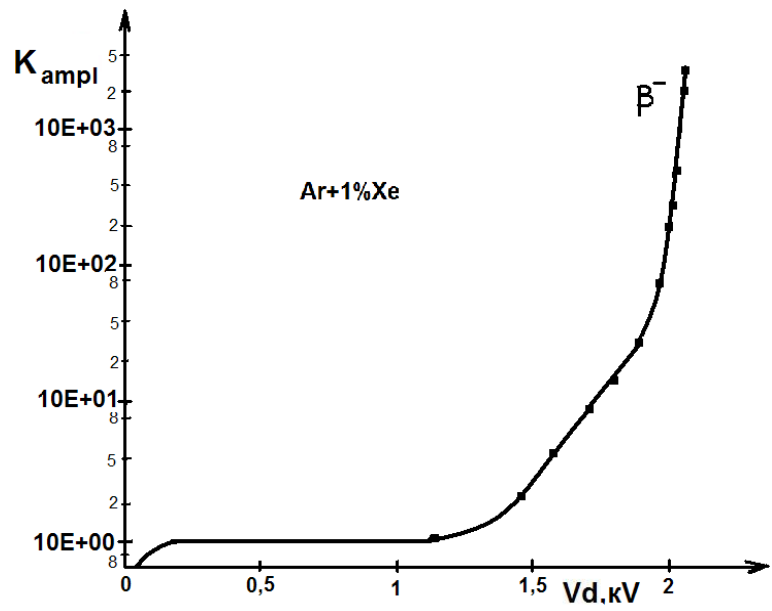

Fig. 9. $K_{ampl}=f(V_d)$ of proportional signals for MWCAT at *Ar+1% Xe* filling.